\documentstyle[aps,prl,epsf,floats]{revtex}

\begin{document}

\ifpreprintsty \else
\twocolumn[\hsize\textwidth\columnwidth\hsize\csname@twocolumnfalse%
\endcsname \fi

\draft
\title{Dynamical frictional phenomena \\ in an incommensurate two-chain
model}
\author{Takaaki Kawaguchi}
\address{Department of Technology, Faculty of Education, Shimane University\\
1060 Nishikawatsu, Matsue 690, Japan}
\author{Hiroshi Matsukawa}
\address{Department of Physics, Osaka University\\
1-16 Machikaneyama Toyonaka, Osaka 560, Japan}
\maketitle
\begin{abstract}

Dynamical frictional phenomena are studied theoretically in a two-chain
model with incommensurate structure.
A perturbation theory with respect to the interchain interaction
reveals the contributions from
phonons excited in each chain to the kinetic frictional force.
The validity of the theory
is verified in the case of weak interaction by comparing
with numerical simulation.
The velocity and the interchain interaction dependences of the lattice
structure are also investigated.
It is shown that peculiar breaking of analyticity states appear,
which is characteristic to the two-chain model.
The range of the parameters in which the two-chain model is
reduced to the Frenkel-Kontorova model is also discussed.

\end{abstract}


\ifpreprintsty \else
] \fi              

%

\section{Introduction}
The sliding velocity dependence of the kinetic frictional force is of
particular importance in the field of tribology. 
The Coulomb-Amonton's law states that the kinetic
frictional force does not depend on the sliding velocity \cite{bow}.
It is well known that this law holds well under usual condition, 
but breaks in some cases.
Its physical
explanation is not well-established. In recent years the study of atomic
scale frictional phenomena has been a hot issue
\cite{binnig,krim1,per,sliding}. In such microscopic systems, new physical 
laws and concepts of friction is considered to exist.
Several works on kinetic friction have been reported using
microscopic or atomic scale models\cite{sliding}.
Among these studies, the Frenkel-Kontorova model (FK model)\cite{fk}
and related ones have been investigated extensively by several
researchers 
\cite{aubry0,aubry1,copper1,copper2,soko1,hira1,shinjo,matsu1,etc,elmer1,matsu2}.
The FK model consists of the atoms interacting each other via harmonic force
under a periodic potential.
In the case that the ratio between the mean atomic spacing and the
period of the potential is irrational, i.e., in an incommensurate case,
the FK model shows a phase transition, the so-called Aubry transition
\cite{aubry0,aubry1}.
When the amplitude of the periodic potential is smaller than a
certain critical value, the maximum static frictional force vanishes.
Above the critical amplitude it becomes finite.
The dynamical properties of the FK and related models have been
reported by many authors.
Coppersmith and Fisher investigated the behavior near 
the maximum static frictional force in the FK
model from the viewpoint of critical phenomena
\cite{copper1,copper2}. Then, the critical exponent of the
velocity-force characteristics was evaluated.
Sokoloff developed a perturbation theory of frictional force
\cite{soko1}. He employed a three-dimensional FK model, which
consists of a three-dimensional deformable lattice and a rigid
substrate, which makes periodic potential.
He discussed frictional effects in commensurate and incommensurate
systems and found that strong reduction of kinetic frictional force
for an incommensurate system compared with that for a commensurate one.
The existence of a superlubricity state, in which the
frictional force vanishes, was claimed by Hirano and Shinjo 
in a three-dimensional FK model \cite{hira1,shinjo}. 
Since they assumed an energy conservative system, 
the kinetic frictional force vanishes by the
energy recurrence effect between the motion of the center of gravity
and the lattice vibration.
Matsukawa and Fukuyama investigated the static and kinetic
frictional forces based on a one-dimensional
model\cite{matsu1,matsu2}. The model employed in their study consists
of two atomic chains, where the interchain interaction, the harmonic
intrachain interaction and the effect of energy dissipation 
are taken into account.
In their model a definitive expression of both of static and kinetic
frictional forces can be derived.
Using a numerical simulation, they found that the critical amplitude
of interchain potential  where the Aubry transition occurs depends
strongly on the elasticity of the chains.
They investigated also the velocity dependence of the kinetic
frictional force.
When the interchain interaction is weak, the crossover behavior of
the kinetic frictional force between velocity-strengthening to
velocity-weakening is observed as velocity increases. It was also 
found that the crossover velocity and the strength of the kinetic
frictional force make obvious difference whether the lower chain is
deformable or rigid. For strong interchain interaction, the
velocity-strengthening behavior is smeared out because of large 
maximum static frictional force.
Quasi-periodic and chaotic sliding
states were discussed by Elmer, Strunz and Weiss 
in the underdamped regime of the FK and FK-Tomlinson models
\cite{elmer1}.

For the theoretical understanding of the dynamics of frictional phenomena,
it is desirable to clarify the behavior and features of kinetic
frictional force both by numerical and analytical methods.
In this paper, using the one-dimensional two-chain model of friction
proposed in Ref. \cite{matsu1}, 
we revisit the dynamic properties 
of the model and calculate the kinetic frictional force by employing both of 
perturbation theory and numerical calculation. 
On the basis of a theoretical expression of frictional force \cite{matsu1}, 
we can successfully formulate the perturbation 
theory on the calculation of the kinetic frictional force for the 
two-chain model.
The role of the phonon excited in each chain can be clarified.
Numerical simulations are employed to
clarify the validity of the perturbation theory and to investigate the 
atomic configuration and motion.
The velocity dependence of the kinetic frictional force under various 
conditions is investigated. 
The physical meaning of the earlier results\cite{matsu1} on the kinetic 
friction is discussed. 
We also investigate the hull function, which characterize the lattice structure. 
It is found that peculiar breaking of analyticity states appear, 
which is characteristic to the case of two deformable chains.
Lastly we discuss the difference between the two-chain model and the FK model and 
the parameter regime in which the two-chain model is reduced to the FK model 
approximately.
\section{Two-chain model of friction}

The two-chain model of friction is reviewed in the following.
We consider two atomic chains, i.e., an upper chain and a lower
chain.
The atoms in each chain have one-dimensional degree of freedom
parallel to the chain and interact each other via the 
harmonic force.
Interchain interaction works between atoms in the upper chain 
and those in the lower one.
The effects of energy dissipation in both chains are assumed to be 
proportional to the difference between the velocity of each atom 
and that of the center of gravity.
The upper chain is subject to the external force parallel to the
chain.
Here we assume overdamped motion, and then the equations of
motion of the atoms in the upper and lower chains are expressed as,
\begin{eqnarray}
m_a\gamma_a(\dot{u}_i-\langle\dot{u}_i\rangle_i)
&=&  K_a(u_{i+1}+u_{i-1}-2u_i) \nonumber \\
&& +\sum^{N_b}_{j\in b} F_I(u_i-v_j) +F_{ex},\\
m_b\gamma_b(\dot{v}_i-\langle\dot{v}_i\rangle_i)
&=&  K_b(v_{i+1}+v_{i-1}-2v_i) \nonumber \\
&& +\sum^{N_a}_{j\in a} F_I(v_i-u_j) - K_s(v_i-ic_b),
\end{eqnarray}
where $u_i$ ($v_i$), $m_a$ ($m_b$), $\gamma_a$ ($\gamma_b$), 
$K_a$ ($K_b$) and $N_a$ ($N_b$) 
are the position of the i-th atom, 
the atomic mass, 
the parameter of energy dissipation, 
the strength of the interatomic force  
and the number of atoms in the upper (lower) chain, respectively.
$K_s$ is 
the strength of the interatomic force between the lower chain and 
the substrate and 
$\langle\dots\rangle_i$ 
represents the average with respect to $i$.
$F_I$ and $F_{ex}$ are the interchain force between the two atomic
chains and the external force, respectively.
We adopt the following interatomic potential:
\begin{equation}
U_{I}=-\frac{K_I}{2}\exp{\left[-4\left(\frac{x}{c_b}\right)^2\right]},
\end{equation}
where $K_I$ is the strength of the interchain potential, $c_b$ the
mean atomic spacing of the lower chain.
The interatomic force is given by $F_I(x)=-\frac{d}{dx}U_{I}$.
The time-averaged total frictional force of the present model 
is given by the following equation.
\begin{equation}
F^{\rm fric}= 
- \sum^{N_a}_{i\in a}\sum^{N_b}_{j\in b}\left\langle F_I( v_i-u_j)
\right\rangle_{t} = N_a \left\langle F_{ex} \right\rangle_{t}
\end{equation}
It should be noted that the above expression of the frictional force is
valid for both static and kinetic ones.
In this study we apply Eq. (4) to the evaluation of kinetic frictional
force.
\section{Perturbation theory of frictional force for two-chain model}

We construct a perturbation theory for the kinetic frictional force
that arises between two interacting atomic chains.
In a steady state the upper chain is assumed to be sliding at a
constant velocity, $V$, on average. Then the atomic positions 
in the upper and
lower chains are expressed as,
\begin{eqnarray}
u_i &=& Vt + ic_a + \delta u_i,\\
v_i &=& ic_b + \delta v_i,
\end{eqnarray}
where $c_a$$(c_b)$ and $\delta u_i$$(\delta v_i)$ denote the mean lattice 
spacing and the deviation of 
atomic position 
from regular site in the upper(lower) chain, respectively.
By substituting them into Eq. (4), we can expand the equation in terms of 
$\delta u_i$ and $\delta v_i$,
\begin{eqnarray}
\lefteqn{
\sum^{N_a}_{i\in a}\sum^{N_b}_{j\in b} \langle F_I(u_i-v_j)
\rangle_{t} }\nonumber\\
&=& \sum^{N_a}_{i\in a}\sum^{N_b}_{j\in b} \langle F_I(Vt + ic_a +
\delta u_i-jc_b - \delta v_j )\rangle_{t}\nonumber\\
&\approx& \sum^{N_a}_{i\in a}\sum^{N_b}_{j\in b} \left[ \langle
F_I(Vt + ic_a -jc_b)\rangle_{t}\right.\nonumber\\
&& +\langle \frac{\partial}{\partial u_i}F_I(u_i-v_j)|_{u_i = Vt +
ic_a,v_j = jc_b}\delta u_i \rangle_{t}\nonumber\\
&& \left.+\langle \frac{\partial}{\partial v_j}F_I(u_i-v_j)|_{u_i =
Vt + ic_a,v_j = jc_b}\delta v_j \rangle_{t}
\right].
\end{eqnarray}
In this equation the first term vanishes because the present
model has an incommensurate lattice structure and then the
summations with respect to $i$ and $j$ cancel out.
The second term represents the contribution from the phonon excited in
the upper chain when the atoms in the lower chain are fixed at the
regular sites.
On the other hand, the third term represents the contribution 
from the phonon excited in the lower chain when the upper chain is 
rigid, but moving at a constant velocity $V$ relative to the lower chain.
Then the frictional force is simply given by the
sum of both the lowest-order terms.
%
Because atomic displacements $\delta u_i$ and $\delta v_i$ are caused by 
the interchain interaction, 
the calculation of the frictional force in the lowest order perturbation, 
Eq. (7), assumes that 
$\delta u_i \neq 0$ and $\delta v_i=0$ in the second term, and 
$\delta u_i=0$ and $\delta v_i \neq 0$ in the third term, respectively. 
The contributions from terms 
in which both $\delta u_i$ and $\delta v_i$ are nonzero arise from 
higher order perturbations. 


First of all, we examine the second term in Eq. (7). It is of
particular interest because the friction in the FK model is given 
by this term as discussed just below.
The force that acts on an atom in the upper chain 
in the leading order of $\delta v_j$ 
is expressed by the following Fourier series.
\begin{eqnarray}
\lefteqn{
\left. \sum^{N_b}_{j\in b} F_I(u_i-v_j) \right|_{v_j=jc_b} }
\nonumber \\
&\approx& -0.47{K_I}\sin\left( 2\pi\frac{u_i}{c_b} \right) 
-0.00057{K_I}\sin\left( 4\pi\frac{u_i}{c_b} \right).
\end{eqnarray}
Hence, we can successfully neglect the second term in Eq. (8).
Then, the equation of motion of the upper chain 
in the present perturbation theory approximately 
corresponds to the following equation of motion.
\begin{eqnarray}
m_a\gamma_a \delta\dot{u}_i  
&=&  K_a\{ \delta u_{i+1}+\delta u_{i-1}-2\delta u_i\} \nonumber \\
&&-{0.47K_I}
\sin\left[ \frac{2\pi}{c_b} (Vt+ ic_a +\delta
u_i)\right] .
\end{eqnarray}
This is nothing but the overdamped equation of motion of the FK model. 

In the perturbation theory \cite{soko1,matsu3}, 
the atomic displacement is expressed by the recursive equation, 
\begin{eqnarray}
\delta u_i &=& \frac{0.47{K_I}}{m_a} \sum^{N_a}_{j} \int dt' G_{ij}(t-t')
\nonumber \\
&& \left\{
\sin\left[ \frac{2\pi}{c_b} (Vt' + jc_a +\delta u_j)\right] 
\right.\nonumber \\
&&\left. -\left\langle
\sin\left[ \frac{2\pi}{c_b} (Vt + ic_a +\delta u_i)\right] 
\right\rangle_{i,t}
\right\},
\end{eqnarray}
where $G_{ij}(t-t')$ is the Green function of phonons defined as
\begin{eqnarray}
G_{ij}(t) &=& \frac{1}{N_a} \sum_{k} \int \frac{d\omega}{2\pi}
G_{k}(\omega) e^{ik(i-j)c_a + i\omega t}, \\
G_{k}(\omega) &=& (i\gamma_a\omega + {\Omega (k)}^2)^{-1}, \\
{\Omega (k)}^2 &=& \frac{2K_a}{m_a}\left( 1-\cos k c_a\right).
\end{eqnarray}
In the lowest order, Eq. (10) is reduced to 
\begin{eqnarray}
\delta u_i \approx 0.47\frac{K_I}{m_a} \sum^{N_a}_{j} \int dt'
G_{ij}(t-t')
\sin\left[ \frac{2\pi}{c_b} (Vt' +j c_a)\right] .
\end{eqnarray}

Then the kinetic frictional force per atom in the upper chain resulting from 
the phonon excitation there, $F^{\rm upper}$, is given in the lowest order of 
$K_I$ as, 
\begin{eqnarray}
F^{\rm upper} &{\approx}&
-0.47{K_I}\frac{2\pi}{c_b}\frac{1}{N_a}\sum_{i}^{N_a}\left\langle
\cos\left[ \frac{2\pi}{c_b} (Vt + i c_a) \right] \delta u_i
\right\rangle_{t}  \nonumber \\
&\approx& \frac{(0.47K_I)^2}{2\gamma_a m_a}\frac{\left(
\frac{2\pi\gamma_a}{c_b}\right)^2 V}
{{\Omega \left(\frac{2\pi}{c_b}\right)}^4+
\left( \frac{2\pi\gamma_a}{c_b} V \right)^2}.
\end{eqnarray}
As easily seen by this equation there exists the crossover behavior 
of the velocity dependence of the kinetic frictional force between 
the velocity-strengthening and velocity-weakening.
This is the essential feature of the kinetic frictional force 
in the case of small interchain interaction 
in the present model, which 
is observed in the numerical simulation in Ref. \cite{matsu1} and in section V.
The crossover velocity is determined by three characteristic quantities: 
the phonon frequency at the wavenumber corresponding to the period of 
the underlying potential, the so-called washboard frequency 
$2\pi V/ c_b$ and the damping constant.
In order to consider the reason of this crossover behavior, 
it is interesting to see the average of the squared velocity fluctuation, 
$\langle {\delta \dot{u}_i(t) }^2 \rangle_{i,t}$ given by, 
\begin{eqnarray}
\langle {\delta \dot{u}_i(t) }^2 \rangle_{i,t}  &=&
2\left(\frac{0.47K_I}{m_a}\right)^2
\frac{\left(\frac{2\pi}{c_b}V\right)^2}
{{\Omega \left(\frac{2\pi}{c_b}\right)}^4+
\left( \frac{2\pi\gamma_a}{c_b} V \right)^2} .
\end{eqnarray}
The squared velocity fluctuation shows velocity-strengthening behavior 
in the low velocity regime and then saturates. 
The former behavior comes from the increase of $V$, itself. 
The latter one results from the suppression of the fluctuation of 
position, $\langle {\delta {u}_i(t) }^2 \rangle_{i,t}$, 
in the large velocity regime, as mentioned below.
We note here that 
the kinetic friction in the present model
is caused by the energy dissipation due to the damping of the phonons 
excited by the sliding motion.
Hence the energy dissipation per unit time in the upper chain, 
$F^{\rm upper}V$, is equal to 
$m_a \gamma_a\langle {\delta \dot{u}_i(t) }^2 \rangle_{i,t}$. 
Then the frictional force, which is the energy dissipation per unit sliding 
distance, is equal to 
$m_a \gamma_a\langle {\delta \dot{u}_i(t) }^2 \rangle_{i,t} /V$.
So the velocity-strengthening behavior of $F^{\rm upper}$ in the low veocity 
regime comes from that of 
$\langle {\delta \dot{u}_i(t) }^2 \rangle_{i,t}$.
As $V$ becomes greater than 
$\frac{c_b}{2\pi \gamma_a}{\Omega \left (\frac{2\pi}{c_b}\right)}^2$, 
$\langle {\delta \dot{u}_i(t) }^2 \rangle_{i,t}$ saturates and then 
the frictional force decreases.

Next we consider the kinetic friction resulting from the phonon 
excitation in the lower chain.
In this case we can express the interatomic force that act on an atom in
the lower chain using a Fourier series as follows.
\begin{eqnarray}
\left. \sum^{N_a}_{i\in a} F_I(u_i-v_j) \right|_{u_i=Vt+ic_a} 
&\approx& - 0.83{K_I}\sin\left(2\pi\frac{v_j-Vt}{c_a} \right) 
\nonumber \\
-0.098{K_I}\sin\left(4\pi\frac{v_j-Vt}{c_a} \right).
\end{eqnarray}
We can neglect the second term.
The strength of interchain interaction is not equal to that in Eq. (8). 
This comes from the spatial profile of the interchain
potential and the summation on the upper lattice site.
Then the equation of motion is given by
\begin{eqnarray}
m_b\gamma_b \delta\dot{v}_i
&=&  K_b\{ \delta v_{i+1}+\delta v_{i-1}-2\delta v_i\} - K_s\delta
v_i \nonumber \\
&& - 0.83K_I \frac{2\pi}{c_a}\cos\left[ \frac{2\pi}{c_a} (-Vt+ ic_b)\right]
\delta v_i.
\end{eqnarray}
After a similar calculation to that of $F^{\rm upper}$, we get
the kinetic frictional force resulting from the phonon excitation 
in the lower chain. It is written as $F^{\rm lower}$ and is given by
\begin{eqnarray}
F^{\rm lower} =
\frac{N_b}{N_a}\frac{\left(0.83K_I\right)^2}{2\gamma_b
m_b}\frac{\left( \frac{2\pi\gamma_b}{c_a}\right)^2 V}
{{\Omega_b \left(\frac{2\pi\gamma_b}{c_a}\right)}^4+
\left( \frac{2\pi\gamma_b}{c_a} V \right)^2},
\end{eqnarray}
where $\displaystyle \Omega_b (k) = 
\sqrt{\left( K_s +2K_b\left( 1-\cos k c_b\right)\right)/m_b}$ is 
the phonon frequency of the lower lattice.
$F^{\rm lower}$ is also evaluated per atom
in the upper chain.

The total kinetic frictional force
$F^{\rm fric}$ is given as 
$F^{\rm fric} = F^{\rm upper}+F^{\rm lower}$.
Then the strength and velocity dependence of the total 
kinetic frictional force are determined by both of 
$F^{\rm upper}$ and $F^{\rm lower}$. 
This is the essential difference with the FK model. 
In the latter $F^{\rm fric}=F^{\rm upper}$, of course.

The perturbation theory is expected to be valid under the 
condition that the atomic displacement of the lattices is
small compared to the mean lattice spacing. That condition 
is expressed as follows.
\begin{eqnarray}
\sqrt{\langle {\delta u_i(t)}^2 \rangle_{i,t} } 
&=&
\frac{0.47K_I}{\sqrt{2}m_a}
\frac{1}{\sqrt{{\Omega \left(\frac{2\pi}{c_b}\right)}^4+
\left( \frac{2\pi\gamma_a}{c_b} V \right)^2}} 
\ll c_a ,
\end{eqnarray}
and
\begin{eqnarray}
\sqrt{\langle {\delta v_i(t)}^2 \rangle_{i,t} } 
&=&
\frac{0.83K_I}{\sqrt{2}m_b}
\frac{1}{\sqrt{ {\Omega_b \left(\frac{2\pi}{c_a}\right)}^4+ \left(
\frac{2\pi\gamma_b}{c_a} V \right)^2}} 
\ll c_b .
\end{eqnarray}
These inequalities depend on velocity because atomic displacements are 
suppressed by large velocity.
The reason is that in the equation of motion, Eqs. (1) and (2), 
the effect of the high-velocity-sliding motion of atoms effectively weakens 
that of interchain interaction, which causes the atomic displacement. 
It is a dynamical effect in sliding.
Here we make some comments on these conditions required for the two chains. 
For the perturbation in the FK model, only one inequality, Eq. (20), is required, 
but in the two-chain model both inequalities must be satisfied simultaneously. 
It is to be noted that in the present lowest order calculation the total kinetic frictional 
force is the sum of the contributions of phonons in the upper and lower chains as mentioned 
above. 
If the conditions Eqs. (20) and (21) are broken, 
there appears the cross term of $\delta u_i$ and $\delta v_i$, 
which expresses the contribution from the interaction of phonons of both chains.

\section{Method of Numerical simulation}

We perform a numerical calculation of the kinetic frictional force and
examine the validity of the perturbation theory developed in the
previous section.
The equations of motion are numerically solved using the Runge-Kutta
formula at the fourth order.
We assume periodic boundary conditions
in both chains. 
Hence the ratio of the mean lattice spacings of the upper and 
lower chains $c_a/c_b$ is equal to the ratio $N_b/N_a$.
We choose the ratio by using the continued-fraction expansion of the
golden mean.
\begin{equation}
\frac{c_a}{c_b}=\frac{N_b}{N_a}=\frac{144}{89}=1.617\cdots,
\end{equation}
where the expansion is truncated at the tenth order.
We set the values of the model parameters as
\begin{eqnarray}
N_a&=&89, N_b=144, c_a=\frac{144}{89}, c_b=1, m_a=m_b=1,\nonumber\\
K_a&=&1,K_b=0, \gamma_a=\gamma_b=1,
\end{eqnarray}
where the intrachain force is neglected, i.e., $K_b=0$, for simplicity. 
Such simplification was employed also in the earlier study \cite{matsu1}.
The lower chain, therefore, consists of individually
oscillating atoms, that is, the Einstein's model of lattice vibration is
adopted in the lower lattice.

Throughout this study, the frictional force is evaluated per atom in
the upper chain.
\section{Comparison between the results of the perturbation theory 
and numerical calculation}

In Section III we have presented the kinetic frictional force calculated 
in the lowest order perturbation. 
Higher order terms of the perturbation can be calculated in a similar manner, but 
it is easier and more straightforward here to check the validity of the lowest order perturbation 
by comparing its result with that of numerical calculation. 
%
First of all, we must investigate the relationship between 
the strength of the interchain interaction and the validity of the perturbation theory 
in the case that the lower chain is fixed, 
which corresponds to the FK model as mentioned in Section III. 
In this case the contribution to the kinetic frictional force results only from the 
phonon excitation in the upper chain.
Even in such a single deformable chain case 
the validity of the perturbation theory has never been examined so far especially 
for strong interchain interaction. 
%
The kinetic frictional force in the case of weak interchain interaction, $K_I=0.1$, 
which is less than the critical value of the Aubry transition, 
is shown in Fig. 1(a) as a function of the sliding velocity.
Then the maximum static frictional force vanishes because of the absence of 
the Aubry transition. 
A good agreement between the result of the lowest order perturbation theory 
and that of the numerical calculation is obtained and 
the velocity-strengthening and velocity-weakening features are
clearly observed.
From this result, it is confirmed that the result of 
the lowest order perturbation theory is quite good. 
This is consistent with the condition, Eq. (20), which gives 
$\sqrt{\langle {\delta u_i(t)}^2 \rangle_{i,t} }\approx 0.01 \ll c_a$ 
even in the case of vanishing velocity.
Hence it is found that dominant contribution to the kinetic frictional force 
comes from the lowest order, and then higher order terms are negligible.
%
On the other hand, for strong interchain interaction, $K_I=1$, which is greater than the 
critical value of the Aubry transition and therefore finite static frictional force 
exists as observed in Fig. 1(b), 
obvious discrepancy is observed between the result of the perturbation theory 
and that of the numerical calculation in the low velocity regime. 
%
This dicrepancy appears to arise from the Aubry transition. 
The effect of the Aubry transition affects seriously the kinetic friction 
in the low velocity sliding state.
However, we should note that 
in the high velocity regime ($V \geq 0.6$) 
the result of the lowest order perturbation and that of the numerical calculation 
agree well.
This is also consistent with Eq. (20). 
The solid lines in Fig. 1 denote $\sqrt{\langle {\delta u_i(t)}^2 \rangle_{i,t} }/ c_a$. 
As seen from Fig. 1(b), the frictional force derived from the lowest order perturbation theory 
agrees well with that from numerical calculation, where the velocity is so large that 
$\sqrt{\langle {\delta u_i(t)}^2 \rangle_{i,t} } \ll c_a$ is satisfied. 
Therefore, in this velocity regime, it is considered that the sliding state 
is well described by the lowest order perturbation theory.
In the following, we analyze the lattice structure of the sliding chain 
and clarify that the discrete features of the atomic distribution caused by 
the Aubry transition is relevant to the above discrepancy in the low velocity regime.
%

In order to show that the validity of the perturbation theory has 
a close relationship to the lattice structure of the sliding chain, 
it would be useful to consider the hull function, $h(x)$. 
It is defined as follows,
\begin{eqnarray}
u_i = i c_a  + h(i c_a ).
\end{eqnarray}
The hull function is a periodic function, the period of which is 
that of the potential: $h(x) = h(x + c_b)$, and therefore the variable
of the hull function is defined in the range from $0$ to $c_b$.
We pay our attention first to the static case ($V=0$). 
When the strength of the interchain interaction is 
less than the critical value of the Aubry transition, 
the hull function is continuous and almost sinusoidal.
This reflects the sinusoidal form of the interchain force of Eq. (8).
Otherwise, however, 
the hull function is discontinuous and has gaps at several positions
because of the breaking of analyticity 
due to the Aubry transition\cite{aubry0,aubry1} as shown in Fig. 2(a).  
In particular, among the gaps, the largest gap is located at  
the half of the period, $c_b/2$.
This largest central gap is considered to characterize the Aubry transition 
and means that the atomic distribution vanishes at the potential maximums 
and the atoms are bound strongly near the potential minimums.
We show here that the hull function gives us an interesting insight  
about the atomic motion in
sliding under the influence of the underlying potential  
as noticed in Ref. \cite{sneddon}.
Using such a dynamical hull function, the atomic position in sliding is 
given by $ u_i = i c_a +Vt + h(i c_a +Vt)$.
Figs. 2(b)$\sim$(e) show some snapshots of the hull function 
at various steady sliding velocities for the strong interchain 
interaction, $K_I=1$. 
When the velocity is small enough (Fig. 2(b)), 
the reminder of the gaps is clearly observed and the shape is 
highly distorted. 
It is due to the large lattice distortion caused by the
Aubry transition, which is responsible for finite static frictional
force.
The lattice still remains highly distorted even in a sliding state.
Such sliding state is characterized by the almost discontinuous and 
nonuniform spatial distribution of atoms which is caused by 
local atomic stick-slip motion 
in which most atoms stay nearby the potential minimums  
and other atoms jump over the potential maximums.
It should be noted that such slip motion, of course, 
needs finite time 
and then the hull function has no gap in the case of finite velocity.
In Fig. 2(b) the reminders of some small gaps are vanishing, 
but that of the largest gap is apparent. 
The deviation of the position of the reminder of the largest gap from 
the half of the period 
is due to the external force added to
the sinusoidal interchain force \cite{fisher}.
Because the effect of 
large lattice distortion due to the Aubry transition is 
not taken into consideration in the present perturbation theory, 
it fails to explain the behavior of the total kinetic frictional force 
$F^{\rm fric}$ obtained by the simulation. 
%
As the velocity increases, the amplitude of the hull function decreases 
and its form becomes smoother 
and comes close to the sinusoidal one.
This means that the large lattice distortion is suppressed 
in the high velocity regime.
In this regime the perturbation theory gives $F^{\rm fric}$ which 
agrees well with that by the simulation.
As mentioned before, the suppression effect can be confirmed directly 
from Eq. (20), 
as shown by the solid lines in Fig. 1.
Even when the lattice structure is subject to the Aubry transition, 
the dynamic suppression effect washes out it in the high velocity regime 
and leads to the sliding state 
that is described by the lowest order perturbation theory. 

It turns out that the frictional property of the single deformable chain 
on a periodic potential can be understood on the basis of the lowest order perturbation 
theory. 
We move on to the two-chain model. 
We clarify the validity of the lowest order perturbation theory also here 
and where is the difference between the two-chain and the FK models. 
Figs. 3(a) and (b) show the kinetic frictional force for $K_s=2$
and $10$ respectively in the case of weak interchain interaction, 
$K_I=0.1$. 
In these cases the maximum static
frictional force is vanishing and the velocity-strengthening and
velocity-weakening features of the kinetic frictional force are
observed.
The contributions to the kinetic frictional force from 
the phonon excitations in the upper and lower chains, 
$F^{\rm upper}$ and $F^{\rm lower}$, 
and the total kinetic frictional force, $F^{\rm fric}$, 
calculated by the perturbation theory 
are also plotted in these figures. 
For $K_s=2$ (Fig. 3(a)), the contribution to the kinetic frictional force 
from the lower chain is dominant in the overall velocity regime.
On the other hand, when the lower chain becomes stiffer, 
$K_s=10$ (Fig. 3(b)), the contributions from both chains become 
important.
Because the crossover velocity between
velocity-strengthening and velocity-weakening features is different
for each chain in this case, the velocity dependence of the total kinetic
frictional force is modified in a velocity regime  
between the two crossover points ($0.7\leq V\leq 4$).
In this velocity regime, the velocity-weakening behavior 
of the kinetic frictional force due to the lower chain
and the velocity-strengthening one due to the upper chain are 
complemental each other, and therefore the total kinetic frictional 
force becomes velocity-insensitive.
It is found in both the cases that the results of the lowest order perturbation 
theory can explain excellently the behavior of the kinetic frictional 
force observed in the numerical simulations 
when the strength of the interchain interaction is less than 
the critical value of the Aubry transition.
Therefore, it is confirmed also in the two-chain model 
that the lowest order perturbation is valid definitively 
in the absence of the Aubry transition, and then higher order contributions 
are quite negligible. 
Thus the lowest order perturbation gives us a correct and 
quite simple viewpoint on the kinetic frictional force of the two-chain 
model in the absence of the Aubry transition 
that the behavior of each chain can be understood as an individual deformable chain 
on the periodic potential and then 
the kinetic frictional force can be also composed of two contributions that come 
from such individual two chains. 
%
If the Aubry transition exists in the two-chain model, such a simple viewpoint on two chains 
is modified largely and altered to a complicated one.
Fig. 3(c) shows the kinetic frictional force for $K_s=2$ 
in the case of strong interchain interaction, $K_I=1$, 
which is the same strength as that in Fig. 1(b).
There exists a obvious discrepancy between the result of the perturbation and that 
of the numerical calculation, but in a high velocity regime a good agreement is obtained. 
Such behavior is quite similar to that for the FK model in Fig. 1(b). 
Then we investigate the difference in the kinetic frictional force and 
the lattice structure 
between the FK model and the two-chain model. 
In particular, it is quite important to understand 
how the strong interchain interaction leads to the Aubry transition in two chains, 
and how both lattices behave in sliding states. 
In order to discuss lattice structures of two chains, 
we define two hull functions for the two chains of the present model 
in the following equations.
\begin{eqnarray}
u_i &=& i c_a +Vt + h_a(i c_a +Vt),\\
v_i &=& i c_b + h_b(i c_b ).
\end{eqnarray}
In this case the periodicity of the hull
functions is expressed as
\begin{eqnarray}
h_a(x) = h_a(x + c_b),\: h_b(x) = h_b(x + c_a).
\end{eqnarray}
The sliding states of the strongly interacting two-chain
model can also be analyzed using these two hull functions. 
Figs. 4(a)$\sim$(e) show the hull functions given by 
snapshots taken in the case of no external force and in steady sliding states 
at several velocities for the system with $K_I=1$. 
For these model parameters, it should be noted that the
Aubry transition occurs in the lower chain in the stationary state 
in the sense mentioned just below. 
It is confirmed from the fact that the hull function $h_b$ in Fig. 4(a) 
shows the largest gap at the half of the period, $c_a/2\approx 0.81$.
On the other hand, the hull function for the upper chain is also much 
affected by the strong interchain interaction and have gap structure. 
But no gap exists at the half of the period, $c_b/2=0.5$.
The state of the upper chain would be a sort of breaking of analyticity 
state, 
but we can distinguish that from the conventional Aubry's breaking of 
analyticity state for the FK model, 
which is characterized by the largest gap at the half of the period.
The interchain force which affects the upper chain is not simple 
and spatially modulated 
because the lower chain is softer in elasticity than the upper one 
and is highly distorted.
The atoms of the soft lower chain 
are bound near the minimum points of the almost periodic potential 
caused by the stiffer upper chain. 
This leads to the conventional 
Aubry's breaking of analyticity state in the lower chain.
Thus, in the stationary state, the elasticities of two chains cause a peculiar 
breaking of analyticity state in each chain when the strength of the interchain 
interaction is greater than a critical value. 
The critical strength of the interchain interaction itself varies with 
the elastic parameters as observed in Ref. \cite{matsu1}. 
These features of the hull functions are quite complicated and the perturbational 
treatment is apparently invalid in this case.

In sliding states with very low velocity, both the hull functions, $h_a$
and $h_b$, retain the reminder of the gap structure and 
are highly distorted (Fig. 4(b)). 
As velocity increases, however, they gradually change their forms, 
approach to sinusoidal ones 
and reduce the amplitudes (Fig. 4(c)).
Such behavior of the hull functions is essentially the same as 
that for the FK model.
The sinusoidal form of the hull function $h_b$ is almost recovered 
at $V\sim 0.658$ as seen in Fig. 4(d), 
but the hull function $h_a$ is still distorted from the sinusoidal form.
This difference in behavior between the two hull functions 
would come from the difference of their elasticities.
For a higher velocity (Fig. 4(e)), both the hull 
functions become sinusoidal obviously.
It is to be noted here that the main contribution to 
the kinetic frictional force 
comes from the lower chain, in which the conventional Aubry transition 
takes place as observed in Fig. 3(c).
Hence the result of the numerical calculation on the kinetic frictional 
force comes closer to that of 
the perturbation theory above the velocity corresponding to Fig. 4(d), 
where $h_b$ recovers sinusoidal form.
Above the velocity corresponding to Fig. 4(e), where both hull functions 
become sinusoidal, they agree quite well.
This means that the result of the numerical calculation 
for the two-chain model agrees well with 
that of the perturbation theory 
only when the effects due to the breaking of analyticity vanish 
in both chains and therefore the sinusoidal forms appear 
in both the hull functions.  
%

Now we consider the conditions required for the validity of 
the lowest perturbation theory, Eqs. (20) and (21).
Fig. 5 shows $\sqrt{ \langle (\delta u_i)^2 \rangle_i}/c_a$ (Eq. (20)) and 
$\sqrt{ \langle (\delta v_i)^2 \rangle_i}/c_b$ (Eq. (21)) as a function of velocity, 
where the parameters are the same as those in Fig. 4.
In the small velocity regime $\sqrt{ \langle (\delta u_i)^2 \rangle_i}/c_a $ 
is quite small ($\approx 0.06$), but $\sqrt{ \langle (\delta v_i)^2 \rangle_i}/c_b$ 
is rather large ($\approx 0.3$). 
The large atomic displacement of the lower chain that is obtained by 
the lowest perturbation theory means that the theory itself is not valid 
although the atomic displacement of the upper chain calculated 
with the theory is quite small. 
In the high velocity regime above the characteristic velocities for two chains, 
which is about unity, the atomic displacements of both chains are suppressed and then 
the lowest order perturbation theory gives well description of the present system.

We discuss here the range of the parameters 
in which the present two-chain model is reduced to 
the FK model, i.e., a model that consists of a one-chain on a rigid periodic potential.
This offers an interesting insight into the two-chain model because 
the importance of being the two chains is understood from it.
In order to investigate the problem, we note the change of the gap structures of 
the hull functions. 
Fig. 6 shows the two hull functions, $h_a$ and $h_b$, in the absence of the external force 
for several $K_s$'s ($K_I=1$ and $K_b=0$).
For small $K_s$ ($K_s=3$, Fig. 6 (a)), as observed also in Fig. 4(a), 
the central gap exists only in $h_b$, i.e., 
the conventional Aubry transition occurs in the lower chain. 
In this case the elasticity of the lower chain has great importance obviously 
because the lower chain has a dominat contribution to the kinetic frictional force.
As $K_s$ is increased ($K_s=5$, Fig. 6 (b)), however, the central gap disappears in $h_b$.
When $K_s$ is increased further ($K_s \geq 10$, Figs. 6 (c) and (d)), 
the central gap appears in $h_a$.
This is obviously contrary to those for small $K_s$'s, but this gap structure of $h_a$ is 
just the same as that shown in Fig. 2 (a) for the Aubry transition in the FK model. 
Then the amplitude of $h_b$ is strongly suppressed and all the gaps shrink. 
In this case the lower chain is very stiff, and its contribution to the frictional force 
is quite smaller than that from the upper chain in the low velocity regime.
In the high velocity regime it is possible that the high-velocity-sliding upper chain excites 
high-frequency phonons in the stiff lower chain, and then a finite contribution to 
the kinetic frictional force from the lower chain is observable 
if the contribution from the upper chain is sufficiently small in this velocity regime.
This is, of course, an essentially different point between the FK and two-chain models.
If our discussion is restricted to the low velocity regime, 
the two-chain model can be reduced to the FK model in the large $K_s$ regime, $K_s \gg 10$ 
for $K_I=1$. 
Then, the maximum static and kinetic frictional forces become close to that 
for the FK model. 
Such behavior of the maximum static frictional force has been reported in Ref. \cite{matsu1}.
The kinetic frictional force in such a case with the same values of the parameters with 
Fig. 3 (c), $K_I=1$ and $K_b=0$, but $K_s$, which is equal to 20, 
is shown in Fig. 7, where the lower chain is very stiff.
The magnitude of the kinetic frictional force in the low velocity regime comes closer to 
that for the FK model in comparison with Fig. 3 (c).
In the high velocity regime, however, the contribution from the stiff lower chain 
nevertheless becomes important.
In Fig. 8 we summarize the characteristic parameter regimes 
for the two-chain model using the parameters, $K_s$ and $K_I$.
In the regime I, $h_a$ shows the largest central gap, 
while there is no central gap in $h_b$, as observed in Figs. 6 (c) and (d). 
That is, as discussed above, the two-chain model is approximated by the FK model 
especially in the stationary and the low-velocity sliding states. 
In the regime II, however, no central gap exists in $h_a$ and the gap structures of 
both the two hull functions become complicated. 
Then the contribution from the lower chain to the frictional force becomes important.
The values of $K_s$ at the boundary between the regimes I and II 
increase linearly with $K_I$. 
It is to be noted that there would be no exact phase boundary between the regimes I and II.

Thus the elasticity of two chains and the interchain interaction for the two-chain model 
affect each other complexly 
when the strength of the interchain interaction is so strong that 
the Aubry transition occurs. 
Then the lowest order perturbation theory is not applicable to two-chain models 
in that case. 
Furthermore the feature characteristic of the two-chain model is more complicated than 
that for the FK model. 
It is found, however, that 
for both models the validity of the lowest order perturbation theory is retrieved 
in the high velocity regime, and 
the two-chain model is reduced to the FK model when the lower chain becomes stiffer than 
the upper chain under a certain condition on the strength of the interchain interaction.
\section{Summary}

We have investigated the kinetic frictional force for a one-dimensional 
two-chain model with an incommensurate lattice structure.
On the basis of the explicit theoretical expression of frictional
force, we have formulated a perturbation 
theory of the kinetic frictional force and clarified the contribution 
from the phonon excitation in each chain to the kinetic frictional force.
From the study on the FK model (single deformable chain model) and the two-chain model 
we have found that the lowest order perturbation 
theory explains the numerical results excellently over a wide range of
velocity if the strength of the interchain interaction is less 
than the critical value of the Aubry transition.
Even if the finite static frictional
force due to the Aubry transition appears, the lowest order perturbation theory 
is still valid in the high velocity regime where the suppression of atomic
displacement becomes significant and the large lattice distortion 
due to the breaking of analyticity is reduced 
by high-velocity sliding motion.
In particular, for the two-chain model with certain elastic parameters, 
we have found that 
the conventional Aubry's breaking of analyticity state, 
which is observed in the FK model and is characterized by 
the largest central gap of the hull function, 
and an another breaking of analyticity state 
appear in the lower and upper chains, respectively. 
The latter lacks the largest central gap and is not well-defined 
in the context of the conventional Aubry transition. 
These stationary states cause complicated sliding lattice structures 
which are responsible for the discrepancy between the result of 
the perturbation theory and that of the numerical simulation. 
As velocity is increased, however, these states of two chains come to 
sliding states described by the lowest order perturbation theory.
Thus, because each chain play an important role on the kinetic friction, 
it is necessary to take into account the contributions to 
the kinetic frictional force from  both chains 
for understanding kinetic frictional force of 
the two-chain model \cite{soko2}. 

We have also clarified the range of the parameters in which the two-chain model is reduced 
to the FK model. 
In the suitable range of $K_I$ and $K_s$ the former reduces to the latter 
in the low-velocity regime. 
Even in that range, however, discrepancy between them appears in the high-velocity regime. 

The present results for the two-chain model is useful to understand 
the earlier results in Refs. \cite{matsu1,matsu2}.
In these studies it was found numerically 
that the kinetic frictional force shows crossover behavior from 
velocity-strengthening to velocity-weakening and the 
crossover velocity and the strength of the kinetic frictional force 
are obviously different between the FK and two-chain models. 
Their results can be understood well with  
the perturbation theory of the present study.

In the present study the parameters of the model are chosen 
artificially and are not necessarily considered the correspondence to 
those in realistic materials because the system is one-dimensional.
Actual frictional phenomena, however,
take place in two- or three-dimensional systems.
The effect of the three-dimensionality of phonon excitation 
and the elasticity in each
substance would be crucial to the kinetic friction in such systems.
Moreover, although the present model has no randomness, 
randomness such as impurities would affect  
the kinetic and static frictional forces. 
The relationship between the velocity dependence of the kinetic 
frictional force and the static frictional force caused by pinning due to 
such randomness is not well established.
These issues will be discussed elsewhere \cite{kawa2}.

\acknowledgments
This work is financially supported by the Sumitomo Foundation and 
Grants-in-Aid for Scientific Research of Ministry of Education, 
Science, Sports and Culture. 


%
%

\noindent
Fig. 1\\
Kinetic frictional force and averaged atomic displacement plotted against 
velocity (for the FK model).
(a) A weak interchain interaction case: $K_I =0.1$. (b) A strong
interchain interaction case: $K_I =1$.
Marked and dotted lines represent the result of the numerical simulation
and that of the perturbation theory, respectively. 
Solid line without marks denotes the averaged atomic displacement 
calculated with Eq. (20), 
which is renormalized by the mean lattice spacing.
\\

\noindent
Fig. 2 \\
Hull functions for the FK model with strong interchain interaction
$K_I =1$. The velocities correspond to the figures are (a) $0$ (no
external force), (b) $2.03\times 10^{-2}$, (c) $1.11\times 10^{-1}$,
(d) $3.95\times 10^{-1}$, and (e) $3.22$.
\\

\noindent
Fig. 3 \\
Kinetic frictional force plotted against velocity (for the two-chain
model).
(a) and (b) correspond to a weak interchain interaction case $K_I
=0.1$, where only $K_s$ is different: (a) $K_s =2$ and (b) $K_s =10$.
(c) is for strong interchain interaction $K_I =1$ and $K_s =2$.
Marked line represents the results of the numerical simulation.
Solid line is the kinetic frictional force obtained by 
the perturbation theory, which is decomposed into 
the contribution from the upper chain (dotted line) and that
from the lower chain (broken line).
\\

\noindent
Fig. 4 \\
Hull functions for the two-chain model with strong interchain
interaction ($K_I =1$) and $K_s=2$. 
The graphs in the left(right) row are the hull functions 
for the upper(lower) chain. 
The velocities are (a) $0$ (no external force), 
(b) $1.36\times 10^{-2}$, (c) $1.52\times 10^{-1}$,
(d) $6.58\times 10^{-1}$, and (e) 1.33.
\\

\noindent
Fig. 5 \\
Averaged atomic displacements plotted against velocity. 
The parameters are the same as those in Fig. 4. 
Solid and dashed lines represent $\sqrt{ \langle (\delta u_i)^2 \rangle_i}/c_a$ 
calculated with Eq. (20) 
and $\sqrt{ \langle (\delta v_i)^2 \rangle_i}/c_b$ calculated with Eq. (21), 
respectively.
\\

\noindent
Fig. 6 \\
Hull functions for the two-chain model with strong interchain interaction 
$K_I =1$. 
$K_s$'s chosen are (a) $3$, (b) $5$, (c) $10$, and (d) $16$.
\\

\noindent
Fig. 7 \\
Kinetic frictional force plotted against velocity.
Marked line with closed squares 
represents the results of the numerical simulation 
for $K_I =1$ and $K_s =20$.
Marked line with open circles 
represents the results of the numerical simulation 
for the FK model with $K_I =1$.
Solid line is the kinetic frictional force obtained by 
the perturbation theory, which is decomposed into 
the contribution from the upper chain (dotted line) and that
from the lower chain (broken line).
\\

\noindent
Fig. 8 \\
Characteristic parameter regimes for the two-chain model.
In the regime I, $h_a$ shows the largest central gap, which is essentially the same as 
that for the FK model, 
and all the gaps of $h_b$ highly shrink and no central gap exists there.
In the regime II, no central gap exists in $h_a$, and both $h_a$ and $h_b$ show 
complicated gap structures.

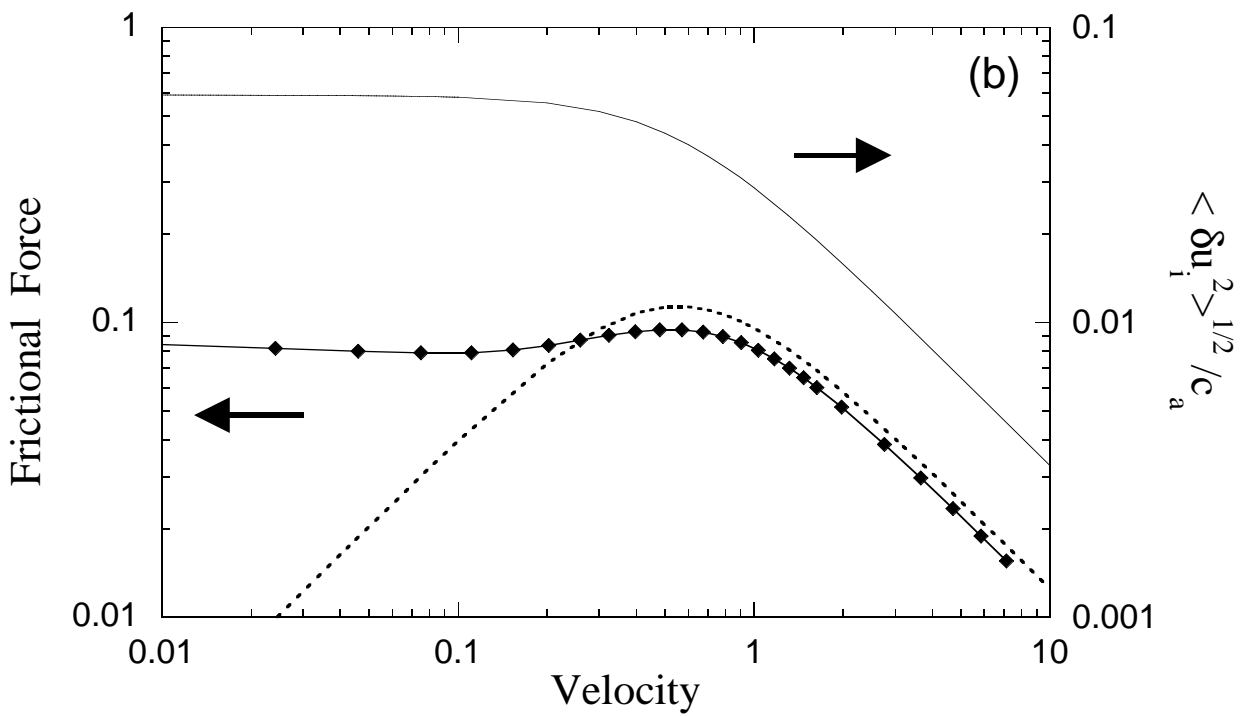
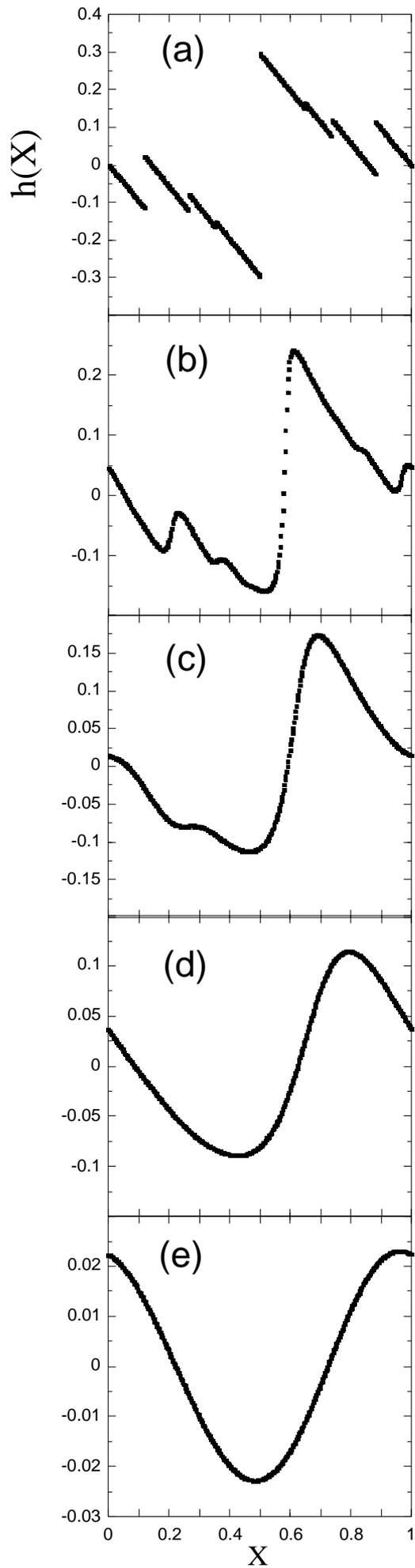
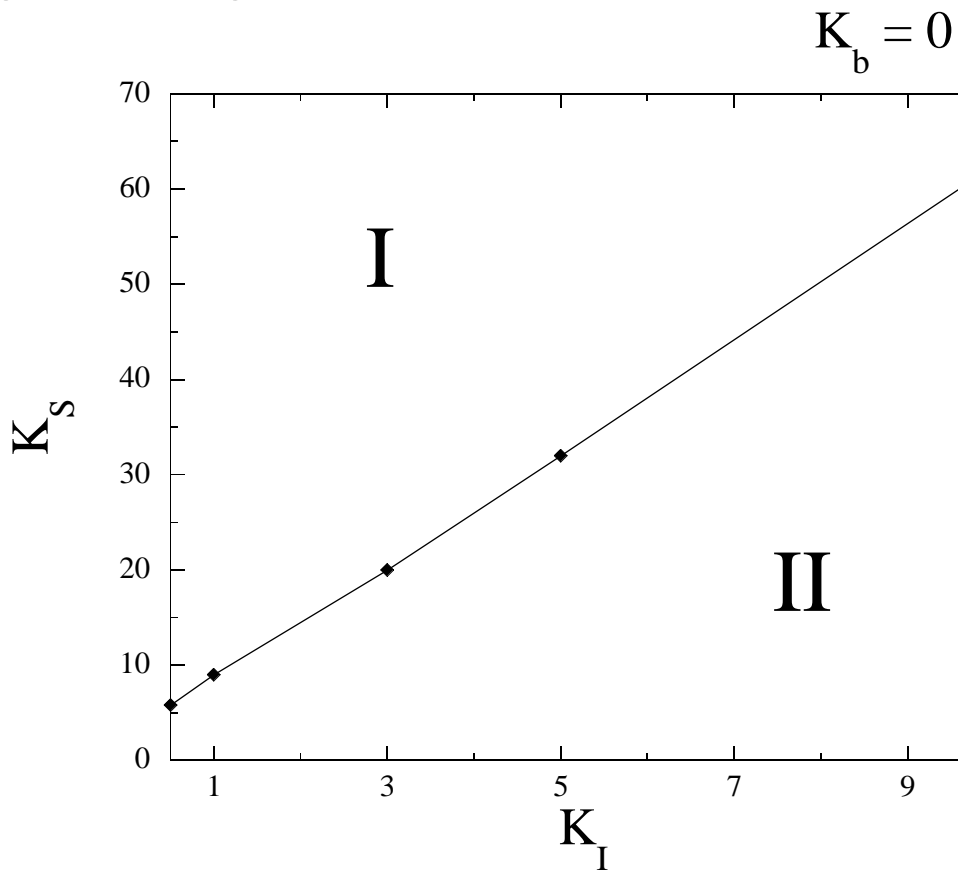
\end{document}